\documentclass{article}

\usepackage{PRIMEarxiv}

\usepackage{booktabs, tabularx}
\usepackage{multirow}
\usepackage{array}
\usepackage{caption}
\usepackage[table]{xcolor} 
\usepackage[utf8]{inputenc} 
\usepackage[T1]{fontenc}    
\usepackage[hidelinks]{hyperref}       
\usepackage{url}            
\usepackage{booktabs}       
\usepackage{amsfonts}       
\usepackage{nicefrac}       
\usepackage{microtype}      
\usepackage{lipsum}
\usepackage{fancyhdr}       
\usepackage{graphicx}       
\graphicspath{{media/}}     

\title{Static Detection of Core Structures in Tigress Virtualization-Based Obfuscation Using an LLVM Pass
}

\author{
  Sangjun An\\
  Chungnam National University \\
  Daejeon\\
  Republic of Korea\\
  \texttt{sangjun0319@gmail.com} \\
   \And
  Seoksu Lee \\
  Chungnam National University \\
  Daejeon\\
  Republic of Korea\\
  \texttt{troy.doubles@o.cnu.ac.kr} \\  
   \And
  Eun-Sun Cho \\
  Chungnam National University \\
  Daejeon\\
  Republic of Korea\\
  \texttt{eschough@cnu.ac.kr} \\
}

\begin{document}
\maketitle

\begin{abstract}
Malware often uses obfuscation to hinder security analysis. Among these techniques, virtualization-based obfuscation is particularly strong because it protects programs by translating original instructions into attacker-defined virtual machine (VM) bytecode, producing long and complex code that is difficult to analyze and deobfuscate. This paper aims to identify the structural components of virtualization-based obfuscation through static analysis. By examining the execution model of obfuscated code, we define and detect the key elements required for deobfuscation—namely the dispatch routine, handler blocks, and the VM region—using LLVM IR. Experimental results show that, in the absence of compiler optimizations, the proposed LLVM Pass successfully detects all core structures across major virtualization options, including switch, direct, and indirect modes.
\end{abstract}


\section{Introduction}
\subsection{Research Background}
As cyber threats become increasingly sophisticated, defense technologies such as malware analysis and reverse engineering are also evolving. Recently, malware has employed obfuscation techniques \cite{collberg1997taxonomy} to evade analysis by security experts. Among these, virtualization-based obfuscation—which protects programs by utilizing a custom Virtual Machine (VM)—is particularly effective. This technique, hereafter referred to as VM-based obfuscation, significantly increases the difficulty of reverse engineering by transforming a program's control flow and data into custom virtual instructions. The resulting VM-obfuscated code is exceptionally large and complex, making it difficult to analyze using conventional methods or AI-based approaches.

Consequently, most current efforts toward devirtualization rely heavily on dynamic analysis, which involves executing the code. However, dynamic analysis faces inherent limitations, such as code coverage issues and the presence of anti-debugging constraints, and even when execution traces are successfully extracted, the sheer volume of the data remains a significant burden.

To address these challenges, this paper proposes a method for analyzing VM-obfuscated code using static analysis, which does not require code execution. We suggest a method to define and extract core structures at the LLVM (Low-Level Virtual Machine) \cite{llvm} Intermediate Representation (IR) level. By developing a static analysis-based tool, we aim to accurately identify the core components of a virtual machine—namely the dispatch routine, handler blocks, and the VM region—to provide stable and versatile information for interpreting the original semantics. While this research focuses on the Tigress obfuscator, it is expected to be extensible to other commercial obfuscation tools in the future. The remainder of this paper is organized as follows: Section 2 introduces the research background and related work. Section 3 analyzes the structure of code protected by VM-based obfuscation. Section 4 presents the methodology for identifying core structures based on static analysis. Section 5 verifies the effectiveness of the proposed static analysis through experimental results, and Section 6 concludes the paper and discusses future work.

\section{Background and Related Work}
\subsection{VM-based Obfuscation}
VM-based obfuscation is an obfuscation technique using virtual machine instructions, making reverse engineering extremely difficult. A virtualized region typically possesses a Fetch-Decode-Execute (FDE) cycle within a switch-loop structure, similar to the instruction processing architecture of a general CPU. It operates by fetching a virtual instruction via a Virtual Program Counter (VPC), decoding it, and then executing the handler corresponding to the identified virtual instruction. Since the meaningful instructions of the program are fragmented within various handlers and the control flow is processed indirectly through a dispatcher, it is very challenging to trace the target code or interpret its actual semantics using conventional analysis techniques.
\begin{figure}[h]
    \centering
    \includegraphics[width=0.4\textwidth]{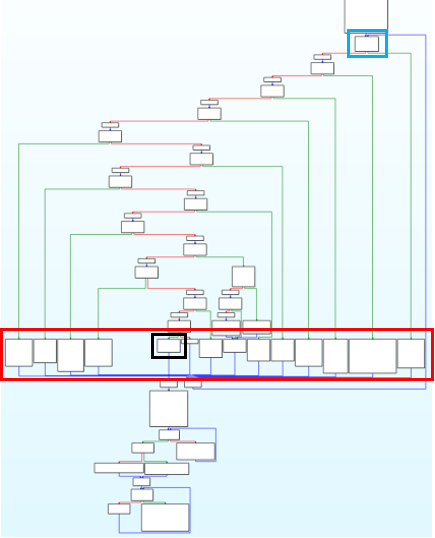} 
    \caption{Control Flow after VM-based obfuscation}
    \label{fig:Fig1}
\end{figure}

Figure \ref{fig:Fig1} shows the Control Flow Graph (CFG) of a code sample where VM-based obfuscation has been applied. The blue box at the top represents the block where the dispatcher begins; it can be observed that the switch structure has been transformed into nested conditional statements through compilation. The blocks within the red boxes are handlers, and among them, the black box in the center is the VM exit handler, which is the block that exits the dispatcher structure.

\subsection{Tigress}
Tigress \cite{tigress}, the target of this research, is a publicly available obfuscation tool. It is a source-code-based obfuscator that takes C code as input and produces obfuscated C code, offering a variety of functions and options such as data obfuscation, control-flow obfuscation, and anti-analysis techniques. The primary virtualization options provided are switch, direct, and indirect. The direct and indirect options represent threaded virtualization designed to conceal the typical switch-loop structure. Among these, the indirect option utilizes table-based branching to specifically increase the difficulty of static analysis.

\subsection{Existing Devirtualization Tools}
To perform devirtualization, it is first necessary to determine whether VM-based obfuscation has been applied. Most existing studies either assume that VM-based obfuscation is already present or utilize AI models to verify its application \cite{yoo2024cnn}. In the simplification stage, dynamic analysis is preferred because of the complexity of VM-obfuscated code, as it allows values to be determined and revealed during execution. In other words, the most common approach involves extracting dynamic traces to analyze virtualization-related structures \cite{salwan, jeon2013dynamic, kim2018code, bang2020vmprotect, park2021analysis, lee2014dynamic, xu2018vmhunt, kalysch2017vmattack}.

However, dynamic analysis for simplifying VM-obfuscated code has several limitations. Above all, if anti-analysis techniques are applied to the obfuscated code, they must be disabled before execution, which imposes a significant burden on the analysis process. Furthermore, there is an inherent limitation regarding the code coverage issue associated with dynamic analysis.
In this study, we propose a method to perform structural analysis based on static analysis using only the code itself, without the need for dynamic traces. Consistent with previous studies that have focused on specific obfuscation tools, this research focuses on the devirtualization of Tigress-obfuscated code.
 
\subsection{LLVM (Low-Level Virtual Machine)}
\begin{figure}[h]
    \centering
    \includegraphics[width=0.6\textwidth]{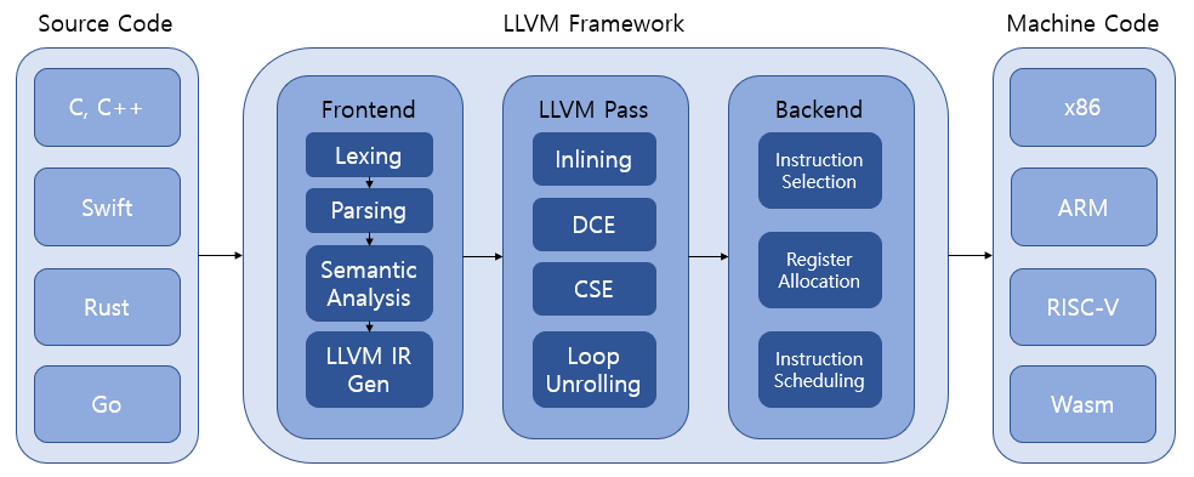}
    \caption{LLVM Framework structure}
    \label{fig:Fig2}
\end{figure}
LLVM, as shown in Figure \ref{fig:Fig2}, is a modular compiler and toolchain framework implemented with separate components: a frontend, an optimizer, and a backend. Consequently, it is an architecture- and language-independent framework that can effectively perform analysis across various languages. In particular, by utilizing LLVM passes, analysts can freely access code blocks, functions, and instructions, allowing proposed analysis techniques to be injected directly into the compilation process.

\section{Analysis of VM-based Obfuscation Structure}
\subsection{Dispatcher Structure Analysis}
In general, the structure of VM-obfuscated code is described by three components: the Virtual Program Counter (VPC), the virtual instruction array, and the dispatcher, which serves as an interpreter to execute them. Among these, the dispatcher acts as the execution framework for the virtual machine based on the VPC, the virtual instruction array, and the handlers. The VPC determines the next handler to be executed, ensuring that the actual instructions corresponding to the obfuscated program are executed in sequence from among the handlers existing in the virtual interpreter array.
\begin{figure}[h]
    \centering
    \includegraphics[width=0.5\textwidth]{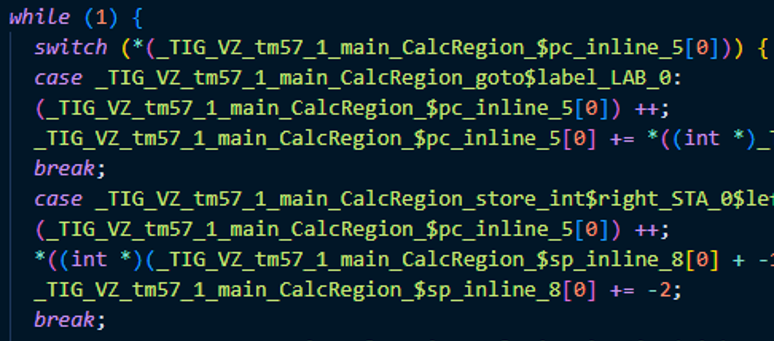} 
    \caption{VM-obfuscated C code with switch option (in part)}
    \label{fig:Fig3}
\end{figure}

The Tigress obfuscation tool provides various dispatch options. In this paper, we analyze the structural changes that occur in the obfuscated code after applying the major options: switch, direct, and indirect. First, the switch option utilizes a switch-loop structure as shown in Figure \ref{fig:Fig3}. Within the loop, the instruction pointed to by the VPC value is selected through a switch-case statement, and the corresponding handler is executed.
\begin{figure}[h]
    \centering
    \includegraphics[width=0.6\textwidth]{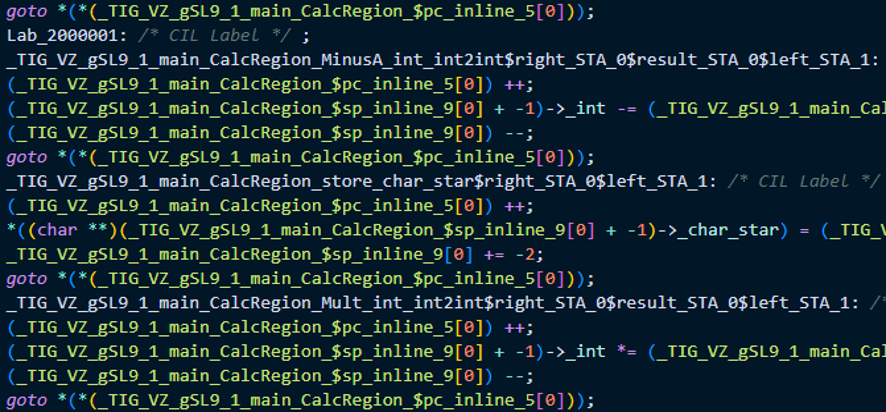} 
    \caption{VM-obfuscated C code with direct option (in part)}
    \label{fig:Fig4}
\end{figure}

The direct option takes the form of threaded code using goto instructions, as shown in Figure \ref{fig:Fig4}. In this threaded code structure, each handler branches directly to the next handler; the VPC is updated within the handler itself, and the starting point of the next handler serves as the destination for the goto statement.
\begin{figure}[h]
    \centering
    \includegraphics[width=0.6\textwidth]{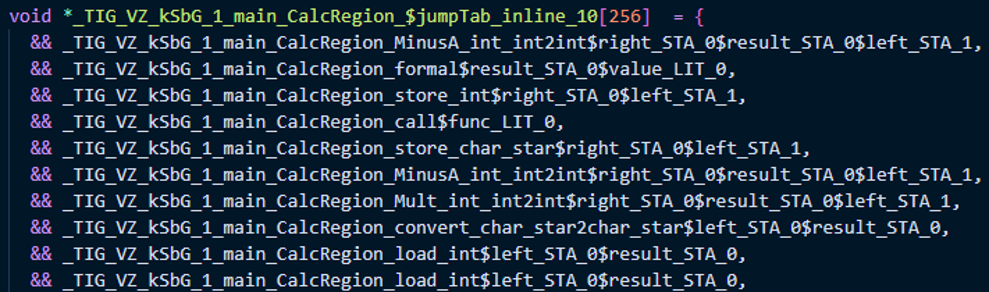} 
    \caption{VM-obfuscated C code with indirect option (in part)}
    \label{fig:Fig5}
\end{figure}

The indirect option is similar to the direct option, but it performs an indirect branch to the corresponding label by referencing a jump table composed of handler addresses using the VPC value at the time of branching (Figure \ref{fig:Fig5}).

\section{Identification of Core Structures Based on Static Analysis}
\subsection{Construction of the LLVM IR Analysis Environment}
To perform static analysis, an environment based on LLVM 22.0.0, Ubuntu 24.04. The experimental dataset was generated by applying VM-based obfuscation to C-based source code using the Tigress 4.0.11 obfuscation tool. To isolate the effects of VM-based obfuscation, other obfuscation were excluded. The Tigress environment was configured for Linux and Clang to ensure compatibility with the execution of the LLVM Pass. Specifically, original C files were obfuscated using specific Tigress options, and the resulting obfuscated C code was converted into LLVM IR via the Clang compiler. Static analysis was then performed using theses IR files. While the current analysis focuses on IR derived from source code, it is expected that this approach can be extended to IR obtained through binary code lifting tools in the future.

\subsection{Design and Implementation of Feature-Based LLVM Pass}
\begin{table}[ht]
\centering
\caption{Description of Core Structures in VM-Obfuscated Code}
\label{tab:Table1}
\renewcommand{\arraystretch}{1.8} 
\begin{tabular}{|>{\centering\arraybackslash}m{3cm}|m{8cm}|} 
\hline
\textbf{Role} & \multicolumn{1}{c|}{\textbf{Description}} \\ \hline
Dispatch Start & The basic block with the highest number of successors within the function. \\ \hline
Handler Block  & All successor blocks of the basic block identified as the Dispatch Start. \\ \hline
VM Start       & The immediate predecessor basic block that calls or branches to the Dispatch Start block. \\ \hline
VM End         & A Handler among the Handler blocks that terminates the program's VM logic instead of branching back to the Dispatch block. \\ \hline
\end{tabular}
\end{table}

Since understanding the architecture of the virtualized machine is crucial for analyzing VM-obfuscated programs, this paper designs and implements an LLVM Pass that detects the core structures of the obfuscated code through LLVM IR analysis. The core structures defined in this study, as shown in Table \ref{tab:Table1}, include the \textit{Dispatch Start}, \textit{Handler Blocks}, and the \textit{VM Start/End}. This approach enables static analysis of the control flow of the obfuscated code without direct execution. Consequently, the specific structural features extracted through static analysis were inserted into the LLVM IR in the form of specially designed dummy function calls, allowing these features to be utilized for core structure detection even at the binary code stage\footnote{The analysis code developed for this study is available on GitHub(https://github.com/sangjun19/LLVM-Pass/tree/main/VMTag)}.

\subsubsection{Identification of the Dispatch Start Point}
Among the virtualization structures, the dispatch start point is the location where the execution framework of the virtual machine begins and serves as a hub block from which branches lead to all handlers.
\begin{figure}[h]
    \centering
    \includegraphics[width=0.5\textwidth]{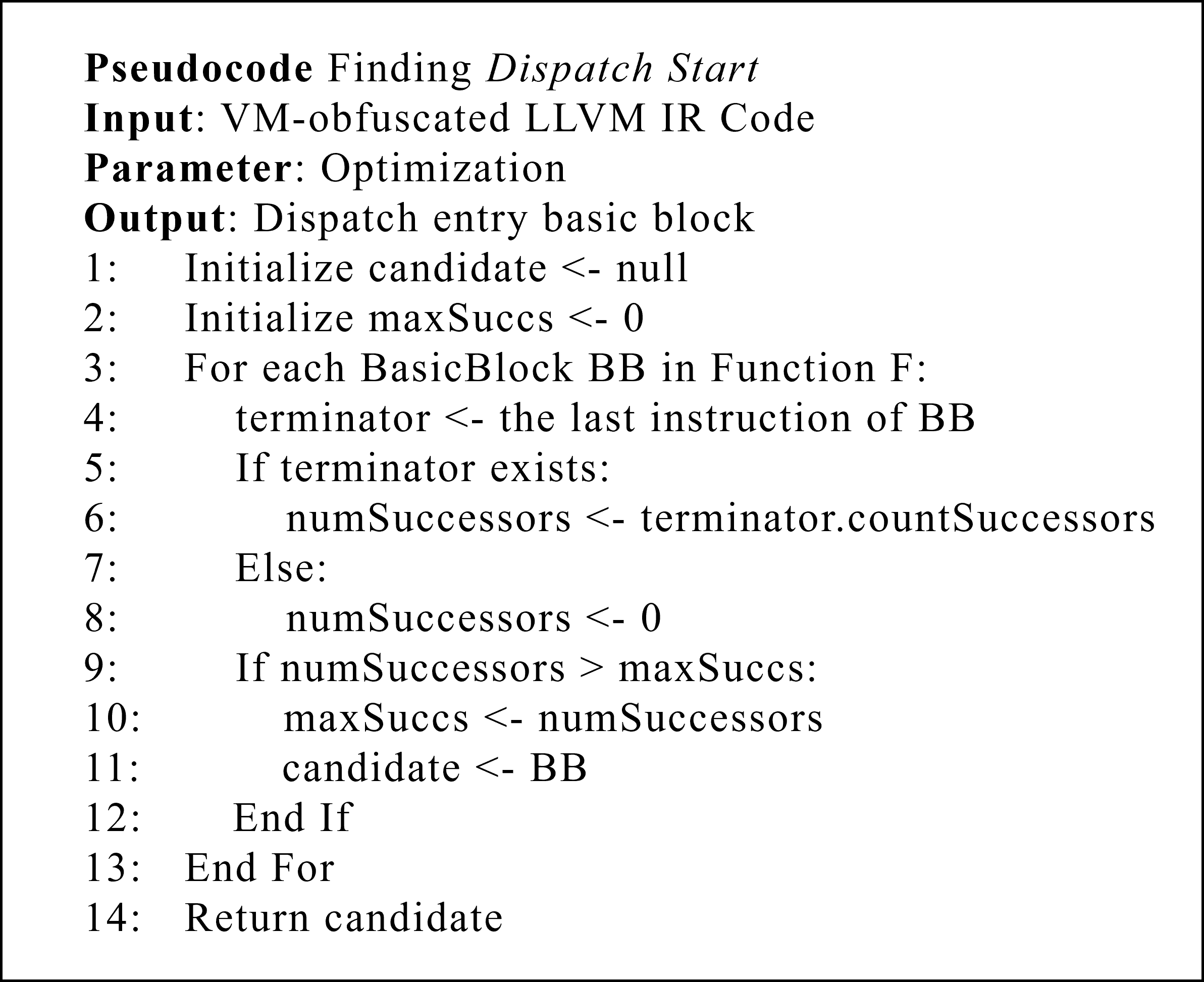} 
    \caption{Locating dispatcher block (LLVM Pass)}
    \label{fig:Fig6}
\end{figure}

Consequently, this \textit{Dispatch Start} point possesses the highest number of exit points (successors). The pseudocode in Figure \ref{fig:Fig6} iterates through all basic blocks within the function and designates the basic block with the highest number of successors as the \textit{Dispatch Start}.
Notably, this central hub block appears even in the direct and indirect options of the Tigress tool, despite their threaded code form. This occurs because, since almost identical goto instructions\footnote{\url{goto *(_TIG_VZ_gSL9_1_main_CalcRegion_$pc_inlinse_5[0]));}} are used at the end of each handler, it was confirmed to be due to the compiler's default behavior of repeatedly utilizing a single common block. Consequently, even in direct and indirect options, the \textit{Dispatch Start} block can be identified by locating the block with the highest number of outgoing edges.

\subsubsection{Identification of Handler Blocks}
A handler block is the smallest unit of code responsible for the actual operation of the original program. Since the previously described \textit{Dispatch Start} serves as the central hub block, all blocks branching from it can be considered handlers. Based on this, we defined all successors of the \textit{Dispatch Start} block as \textit{Handler Blocks}.

\subsubsection{Identification of VM Start/End Points}
The VM region represents the entire scope of the code where VM-based obfuscation is actually applied. \textit{VM Start} is defined as the block that calls or branches to the \textit{Dispatch Start}. We confirmed that this part is indeed the starting point of the VM region by comparing it with the original code in the actual VM-obfuscated code. The \textit{VM End} point is defined as a handler block that terminates the VM logic of the program instead of branching back to the \textit{Dispatch Start} among all \textit{Handler Blocks}.

\section{Experiment}
\subsection{Experimental Environment and Data}
The experimental setup was identical to the previously described LLVM IR analysis environment. The evaluation was conducted using C-based source code featuring various iterative structures, such as sorting algorithms, factorial calculations, and Fibonacci sequences. The control structures of these codes include nested loops, single loops, and single loops within conditional statements.

We applied three major virtualization options (switch, direct, and indirect) of the Tigress obfuscation tool to these C codes. The obfuscated codes were then compiled using the Clang compiler under various optimization levels: no optimization (-O0) and several optimization options (-O1, -O2, -O3, -Ofast, -Og, -Os, -Oz). Through this process, we generated the corresponding LLVM IR code.

In the optimization groups focused on performance and size (-O1 to -Oz), a common simplification of the Control Flow Graph (CFG) was observed. Most notably, there were significant structural changes where basic blocks with a single branch merged with their adjacent blocks. In this study, we designated -O3—the level where these structural transformations occur most aggressively—as the representative optimization group for comparison and analysis against the non-optimized (-O0) environment.

\subsection{Experimental Results and Discussion}
\begin{figure}[h]
    \centering
    \includegraphics[width=0.6\textwidth]{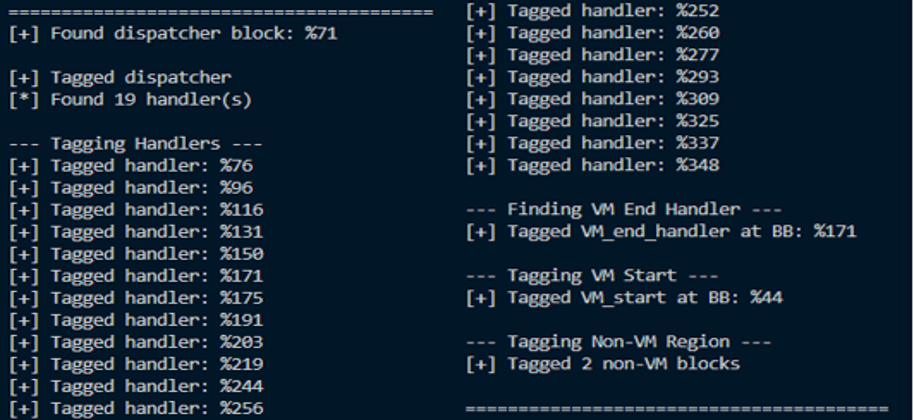} 
    \caption{Execution of the proposed LLVM Pass}
    \label{fig:Fig7}
\end{figure}

\begin{table}[ht]
\centering
\caption{Experimental Results of Core Structure Identification}
\label{tab:Table2}
\renewcommand{\arraystretch}{1.4} 
\small 
\begin{tabular}{|c|c|c|c|c|c|c|c|c|c|c|c|c|c|}
\hline
\multirow{2}{*}{\begin{tabular}[c]{@{}c@{}}\textbf{Opt.}\\ \textbf{Option}\end{tabular}} & \multirow{2}{*}{\begin{tabular}[c]{@{}c@{}}\textbf{Dispatch}\\ \textbf{Option}\end{tabular}} & \multicolumn{4}{c|}{\textbf{Bubble Sort}} & \multicolumn{4}{c|}{\textbf{Factorial}} & \multicolumn{4}{c|}{\textbf{Fibonacci}} \\ \cline{3-14} 
 &  & \begin{tabular}[c]{@{}c@{}}VM\\ Start\end{tabular} & \!Disp.\! & \!Hand.\! & \begin{tabular}[c]{@{}c@{}}VM\\ End\end{tabular} & \begin{tabular}[c]{@{}c@{}}VM\\ Start\end{tabular} & \!Disp.\! & \!Hand.\! & \begin{tabular}[c]{@{}c@{}}VM\\ End\end{tabular} & \begin{tabular}[c]{@{}c@{}}VM\\ Start\end{tabular} & \!Disp.\! & \!Hand.\! & \begin{tabular}[c]{@{}c@{}}VM\\ End\end{tabular} \\ \hline
\multirow{3}{*}{\textbf{-O0}} & switch & O & O & O & O & O & O & O & O & O & O & O & O \\ \cline{2-14} 
 & direct & O & O & O & O & O & O & O & O & O & O & O & O \\ \cline{2-14} 
 & indirect & O & O & O & O & O & O & O & O & O & O & O & O \\ \hline
\multirow{3}{*}{\textbf{-O3}} & switch & X & O & O & X & X & O & O & X & X & O & O & X \\ \cline{2-14} 
 & direct & X & O & O & X & X & O & O & X & X & O & O & X \\ \cline{2-14} 
 & indirect & X & O & O & X & X & O & O & X & X & O & O & X \\ \hline
\end{tabular}
\end{table}

Figure \ref{fig:Fig7} illustrates the results of executing the proposed LLVM Pass on the VM-obfuscated code, demonstrating its ability to identify key structures. Table \ref{tab:Table2} summarizes the detection status of these core structures across the three control structures. In the -O0 environment, our proposed method successfully identified all four core structures within the obfuscated intermediate language: the VM Dispatcher, VM Handlers, and the VM Start and End points.

However, when analyzing LLVM IR with the -O3 optimization applied, we encountered certain limitations in specifying the VM Start block, which is the entry point of the virtualization structure. This occurred because the aggressive optimization of -O3 physically merged the entry of the main function with the VM Start block, causing the two structures to coexist. The \textit{VM End} point exhibited a similar pattern. While the \textit{VM End} in the -O0 environment was an independent block that branched out of the VM region without additional operations, applying optimization caused it to merge with subsequent code outside the VM region.

These merging issues could potentially be resolved by utilizing an LLVM Pass to logically re-separate the structures—for instance, by artificially inserting branch instructions within the merged blocks at the LLVM IR level.

Unlike previous studies, this research distinguishes itself by utilizing static analysis exclusively, without relying on dynamic information. Furthermore, while the current approach targets LLVM IR, it is expected that this technique can be effectively extended to the binary domain once high performance binary to IR liters become available in the future.

\section{Conclusion}
This paper researched a method for statically identifying the core structures of VM-obfuscated code. Through this approach, it is possible to confirm the dispatcher and the core structure of the VM by automatically analyzing only the obfuscated code without the need to extract dynamic traces. We utilized an LLVM Pass and analyzed various virtualization options. In the future, based on these core structural analysis methods, we plan to study the code to perform semantic interpretation of obfuscation and to identify related analyses and roles.

\section*{Acknowledgments}
An extended version of this work has been submitted to the Review of the Korea Institute of Information Security and Cryptology (KIISC).

\bibliographystyle{unsrt}  
\bibliography{references}

\end{document}